\documentclass[12pt,a4paper,final]{iopart}

\usepackage{iopams} 
 \usepackage{paralist}
  \usepackage[english]{babel}
   \usepackage{subcaption}
\usepackage{graphicx}
\usepackage[breaklinks=true,colorlinks=true,linkcolor=blue,urlcolor=blue,citecolor=blue]{hyperref}
\usepackage{breakurl}
\begin{document}

\title[Surface Modifications and Disinfection]{Surface Modifications Caused by Cold Atmospheric Plasma Sterilization Treatment}

\author{Sandra Moritz$^{1,3}$, Alisa Schmidt$^{1,3}$, Joachim Sann$^{2,3}$, M. H. Thoma$^{1,3}$
}
\address{$^1$Justus Liebig University, I. Physical Institute, 35390 Gie{\ss}en, Germany}
\address{$^2$Justus Liebig University, Physical Chemistry Institute, 35390 Gie{\ss}en, Germany}
\address{$^3$Justus Liebig University, Center for Materials Research (LaMa), 35390 Gie{\ss}en, Germany}
\ead{\mailto{sandra.moritz@physik.uni-giessen.de}}

\begin{abstract}
Inactivation of microorganisms on sensitive surfaces by cold atmospheric plasma (CAP) is one major application in the field of plasma medicine because it provides a simple and effective way to sterilize heat-sensitive materials. Therefore, one has to know whether plasma treatment affects the treated surfaces, and thus causes long-term surface modifications. In this contribution, the effect of cold atmospheric Surface Micro-Discharge (SMD) plasma on different materials and its sporicidal behavior was investigated.\\
\noindent Hence, different material samples (stainless steel, different polymers and glass) were plasma-treated for 16 hours, simulating multiple plasma treatments using an SMD plasma device. Afterwards, the material samples were analyzed using surface analysis methods such as laser microscopy, contact angle measurements and X-ray photo-electron spectroscopy (XPS). Furthermore, the device was used to investigate the behavior of \textit{Bacillus atrophaeus} endospores inoculated on material samples at different treatment times.\\
\noindent The interaction results for plasma-treated endospores show, that a log reduction of the spore count between 4.3 and 6.2 can be achieved within 15 min of plasma treatment. Besides, the surface analysis revealed, that there were three different types of reactions the probed materials showed to plasma treatment, ranging from no changes to shifts of the materials' free surface energies and oxidation.\\
\noindent As a consequence, it should be taken into account that even though cold atmospheric plasma treatment is a non-thermal method to inactivate microorganisms on heat-sensitive materials, it still affects surface properties of the treated materials. Therefore, the focus of future work must be a further classification of plasma-caused material modifications.
\end{abstract}
\noindent{\it Keywords}: Cold atmospheric plasma $\cdot$ Spore inactivation $\cdot$ Disinfection $\cdot$ Surface Modification $\cdot$ Surface Micro-Discharge plasma $\cdot$ Plasma surface treatment $\cdot$ Air dielectric barrier discharge $\cdot$ Atmospheric pressure plasma\\
\section{Introduction}
The increase in bacterial resistances causes thousands of deaths every year due to a lack of quick and non-harming sterilization methods. Cold atmospheric plasma (CAP) treatment as a new approach to sterilize surfaces, particularly of heat-sensitive materials or difficult geometries (e.g. \cite{obst,tissue,Sandra,zahn1,zahn2}), is a promising candidate to replace or support conventional cleaning methods.\\
\noindent Although in the last decade cold atmospheric plasma was increasingly used to inactivate bacteria, spores and fungi on different surface materials (\cite{sterilization,spores,infection}), the influence of plasma treatment on these materials was rather neglected. Nevertheless, cold atmospheric plasma is composed of different highly reactive oxygen and nitrogen species (RONS) \cite{Tschang,rons}, e.g. $O_3$, $H_2O_2$, $OH^-$, $HNO_3$ and $N_xO_y$. Hence the reaction between plasma species and surface material has to be well-understood to be able to provide a safe and enduring surface cleaning method.\\
\noindent Recent studies observed etching effects during a dielectric barrier discharge (DBD) treatment \cite{etching}, and oxidation effects of aluminum after 90 minutes treatment with surface micro-discharge (SMD) plasma \cite{aluminium}.  Furthermore, the influence on polytetrafluorethylen (PTFE) and silicon was investigated by Shimizu \textit{et al.} \cite{aluminium}. Nevertheless, due to different setup geometries, plasma production methods, e.g. plasma jets (used for example by Walsh \textit{et al.} \cite{walsh}), Floating-Electrode DBD (e.g. Fridman \textit{et al.} \cite{fridman} or Kuzminova \textit{et al.} \cite{etching}) or SMD (e.g. used by Morfill \textit{et al.}, Kl\"ampfl \textit{et al.} and Shimizu \textit{et al.} \cite{morfill,klaempfl,aluminium}, and different carrier gases (e.g. argon, air and helium \cite{gas,morfill,Sandra,hybrid,helium1,helium2}), the interaction between plasma and target, as well as the gas composition are affected in different ways, and therefore the results are not comparable. Another point that should be taken into consideration is that usually only the effect of short plasma treatments with a time range of 3 to 90 minutes is investigated. Still, sterilization methods for clinical use are performed regularly, often several times per day.\\
\noindent Hence, in this study different material types, regularly used in hospitals, (stainless steel, polymers, glass) were treated with the same plasma source to obtain comparable results. Moreover, treatment time was extended up to 16 hours to achieve clinical-relevant results. Subsequently, laser microscopy, contact angle measurements and XPS quantification were used to investigate surface modifications. As plasma setup, rather than using plasma jet or DBD technology, where the target is in direct contact with the plasma and the produced gas components, an SMD device was used as shown in figure \ref{Aufbau}. Thus, only long-living reactive species of the afterglow, that are more gentle to the material, could interact with the target \cite{morfill,afterglow,afterglowvorteil}. Additionally, the sporicidal properties of the plasma setup were verified probing \textit{Bacillus atrophaeus} endospores on material samples of each group. \\
\section{Materials and Methods}
\subsection{SMD Plasma System}
\begin{figure*}[htb!]
        \centering
               \includegraphics[width=0.5\columnwidth]{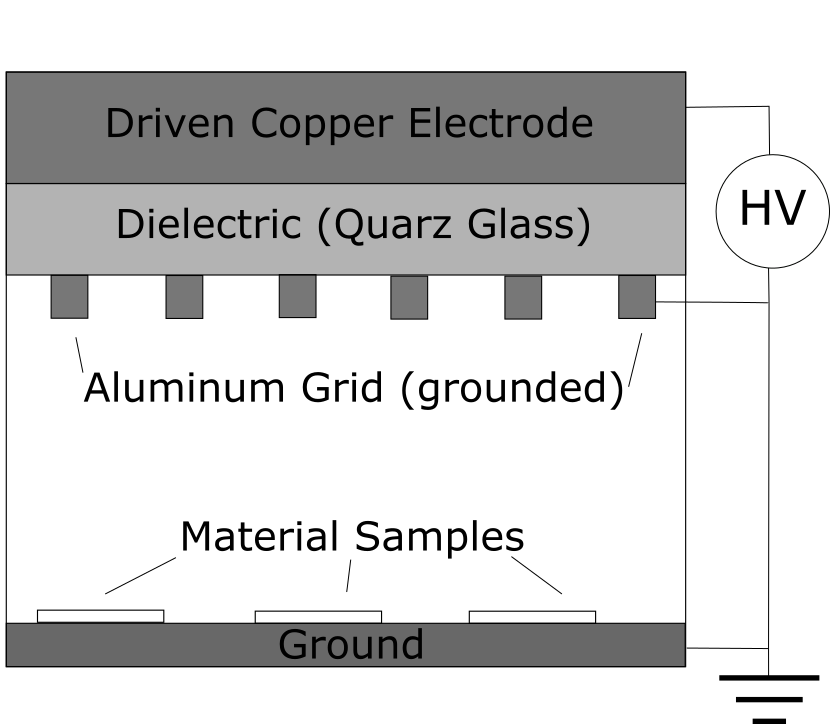}
            
        \caption{SMD plasma setup and electric circuit. A copper electrode was connected to a voltage amplifier, quartz glass was used as dielectric, and a grounded aluminum grid served as counter electrode. Samples were placed at a distance of about 26 mm from the generated plasma on a grounded metal plate.}\label{Aufbau}
  \end{figure*}
 A sketch of the used experimental setup is shown in figure \ref{Aufbau}. The SMD plasma system consists of a copper electrode driven at high voltage, quartz glass as dielectric and a grounded aluminum grid as counter electrode. Within a distance of about 26 mm samples were placed on a grounded metal plate. In this experimental system, the plasma chamber is closed using a movable cover to enable quick air exchange in between plasma treatments.\\
Between grid and copper electrode, high voltage of 9 kV$_{pp}$ at 2 kHz was applied using a high voltage amplifier (model 10/40A, Trek Inc.). SMD plasma is then produced on the surface of the grid electrode. Power density $\left(\frac{P}{A}\right)$ at the SMD surface was measured to be 0.13 $\frac{W}{cm^2}$ using Lissajour's curve \cite{powerdensity}. \\
As material samples, stainless steel AISI 321 (S/S), rigid polyvinyl chloride (UPVC), polypropylene (PP), fluorinated ethylene propylene (FEP) (all GoodFellow) and borosilicate glass (Marienfeld) were chosen, for they are standard materials in the hygiene sector. Sporicidal efficacy of this SMD treatment on the chosen material samples was evaluated by placing the samples (cut to 10 mm x 10 mm; thickness of 0.914 mm (S/S) and 1 mm, respectively; glass samples with a diameter of 10mm and thickness of 0.17 mm) inoculated with \textit{Bacillus atrophaeus} endospores (Simicon) on the grounded metal plate of the plasma setup, and by treating them with afterglow plasma for different time spans, ranging from 5 to 15 minutes.
 For evaluating surface modifications caused by plasma treatment, the plasma chamber containing the samples was driven at room temperature for 16 hours (equivalent to about 190 5-minutes-treatments) with 10 minutes-breaks every 30 minutes. During breaks, the cover was opened to obtain a quick gas exchange with the surrounding air. During all treatments the relative humidity of the ambient environment was about 60 - 70 \%. Note, that S/S samples for surface analysis were mechanically polished first, to improve laser microscopy images. 
\noindent Due to the distance between SMD plasma and sample, the produced plasma species were not instantly filling the whole plasma chamber. Ozone concentration, which was found to be strongly related to bacteria reduction \cite{Sandra, sakiyama}, was measured by absorption spectroscopy (HR4000CG-UV-NIR, Ocean Optics) using collimators, UV optical fibers and light from a UV deuterium lamp at a wavelength of 254 nm (D2000, Mikropack) within the SMD chamber's volume of 429 cm$^3$ at 22 mm distance from the plasma grid, to simulate the height of a sample in a petri dish. The light absorption was measured thrice for 30 minutes with 10 minutes of air exchange between plasma chamber and ambient air in between the measurements. The ozone concentration then was calculated using the Beer-Lambert law with an ozone absorption cross section of 1.1 $\cdot$ 10$^-21$ m$^2$ \cite{beer}.
\subsection{Microbiological Samples}
\begin{figure}[htb!]
 \begin{subfigure}[b]{0.49\columnwidth}
        \centering
           \includegraphics[width=0.9\columnwidth]{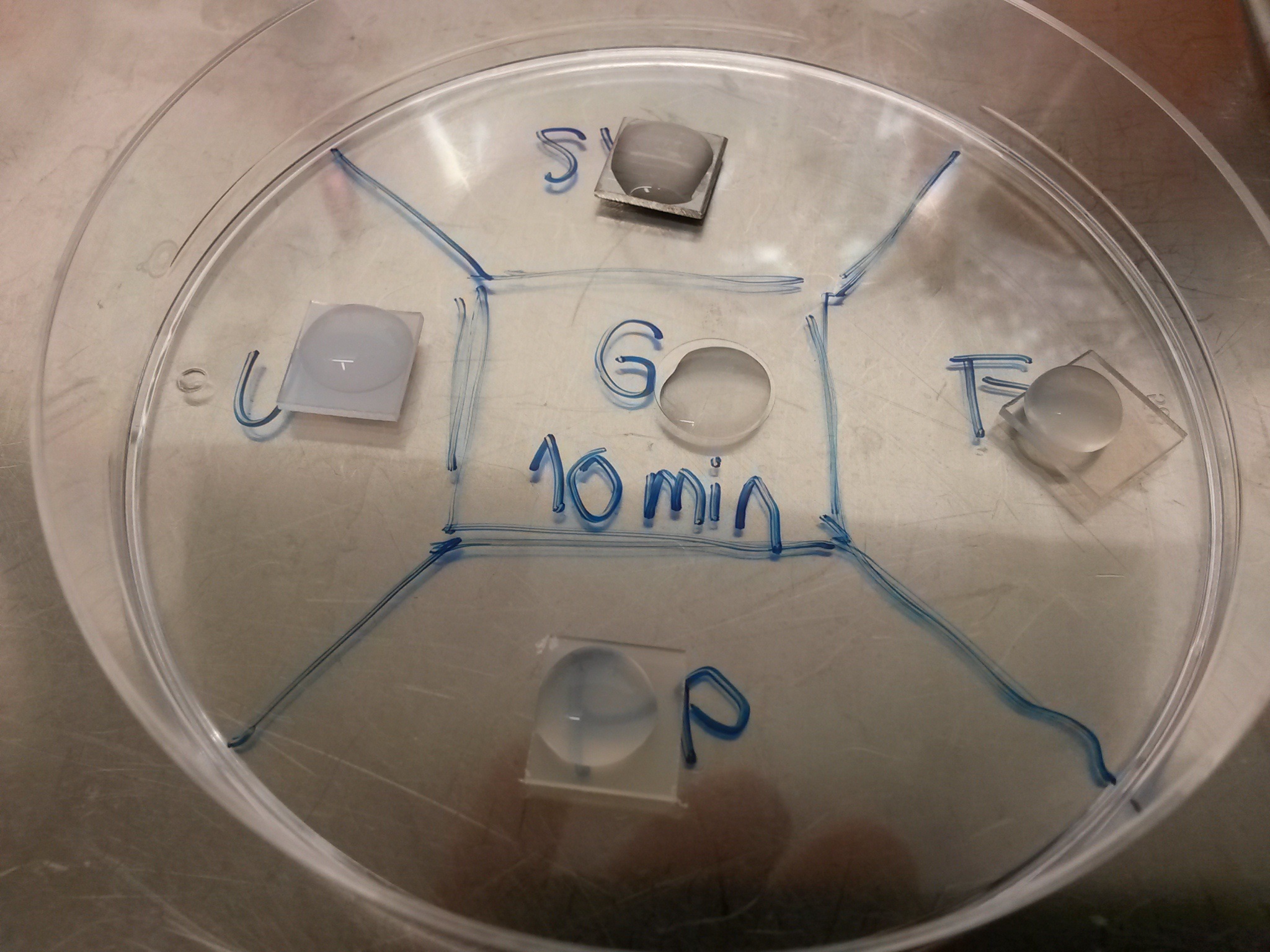}
           \caption{}
           \label{wet}
       \end{subfigure}%
        ~ 
        \begin{subfigure}[b]{0.49\columnwidth}
                \centering
          \includegraphics[width=0.9\columnwidth]{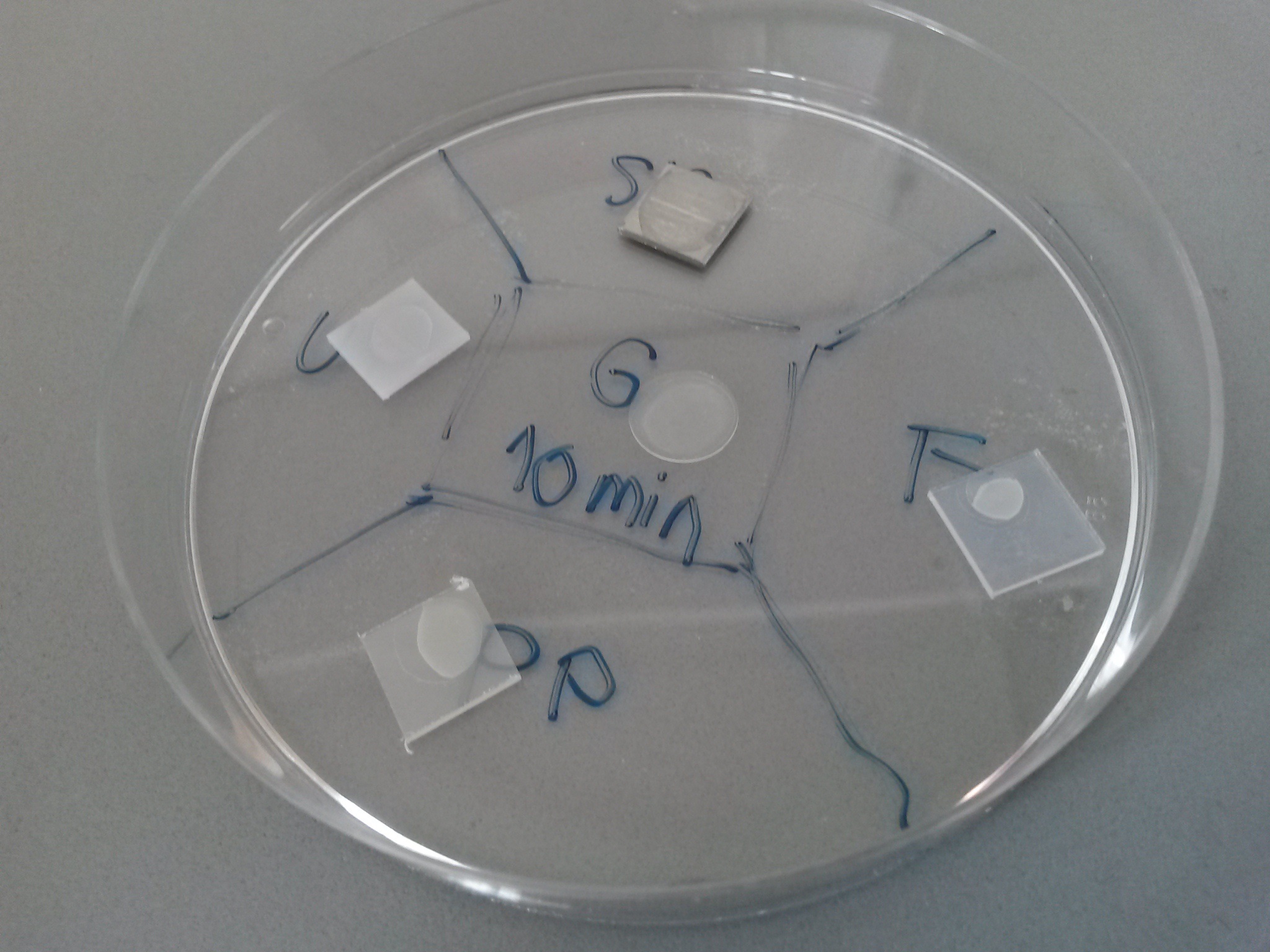}
          \caption{}
                \label{dry}
        \end{subfigure}
        \caption{Material samples inoculated with master suspension of \textit{Bacillus atrophaeus} endospores (left) and after overnight drying process (right).} 
  \end{figure}

\noindent Preparation of spore samples, plasma treatment and recovery of endospores were conducted analogously to the procedure in previous work \cite{Sandra} as listed below:\\
\begin{enumerate}[1.]
    \item As master suspension, \textit{Bacillus atrophaeus} endospores in deionized water were used with a bacteria density of approximately 10$^8$/ml.
    \item 100 $\mu l$ (50 $\mu l$ for glass samples due to smaller geometry) of this master suspension was subsequently spread on the material samples as shown in figure \ref{wet} and dried over night at the ambient air under the safety workbench (\ref{dry}). Note, that the shape of the dried drop differs between different materials, depending on their hydrophilic properties.
    \item The following day, the samples were placed in the plasma chamber as shown in figure \ref{Aufbau}. 
    \item The samples were plasma-treated for 0 (reference samples), 5, 10, and 15 minutes, respectively.
    \item After plasma treatment, the samples were added to 2 ml of deionized water and vortexed for 1 min. This was followed by a 2 min ultrasonic treatment and another vortexing period of 1 min. 
    \item To evaluate the number of surviving endospores, a dilution series was conducted. 100 $\mu l$ of every degree of dilution was spread on agar plates. For 15 minutes treatment the 2 $ml$ of the recovered spore suspension was spread completely on 4 agar plates with 500 $\mu l$ on each.
    \item The agar plates were then incubated for up to 48 hours at 32$^\circ$C.
    \item After incubation the number of colony forming units (CFU) was counted to evaluate log reduction.
\end{enumerate} 
Experiments were performed at least thrice. After experiments, the samples were cleaned with ethanol and autoclaved for 15 minutes at 121$^\circ$ degrees before reusing them.\\
For statistical analysis, diluted samples were only counted, if they had more than 3 CFU. Non-diluted samples with no CFU were counted as if one CFU in the 2 ml suspension was found. In figure \ref{Bakterien} (results), data points, where at least once no CFU was found, are marked with a *.\\
\noindent For our experimental setup, the detection limit in log reduction of \textit{Bacillus atrophaeus} endospores was between 6.2 and 6.8 for the different materials.\\
\subsection{Surface Analysis Methods}
The plasma-treated material samples (as described in the previous section) were tested for surface modifications using laser microscopy, contact angle measurements and XPS quantification. For preventing false results due to material changes caused by surface analysis methods, there was a set of material samples for each analysis method. Recovering effects of the materials, that may occur after plasma treatment, were excluded by performing surface analysis at least 4 weeks after plasma treatment. Reference samples, that were not plasma-treated, were tested at the same time as treated samples.\\
Firstly, for laser microscopy scanning the VK-9710 (Keyence) microscope, using 408 nm wavelength and 50x-fold magnification, provided a topographic scan of the surface, which was then examined using VK Analyzer (Keyence) to specify the roughness of the materials' surfaces. Three samples of each plasma-treated and non-treated material were investigated.\\
Secondly, contact angle measurements were executed using OCA 20 optical contact angle measuring and contour analysis system and SCA 20 and SCA 21 software (DataPhysics Instruments GmbH). As reference fluids, deionized water, ethanol, diiodomethane and ethylene glycol were used. Sessile Drop method \cite{SessileDrop} and, resulting from this data, calculation of the free surface energies of the sample materials were performed, evaluating also the polar and disperse fractions of the surface energy. The reference free surface energies for ethanol were taken from Str\"om \textit{et al.} \cite{strom}, the others from Ohm \textit{et al.} \cite{ohm} to calculate free surface energies of the material samples using OWRK-model \cite{o,r,k}. For every material two plasma-treated samples and two reference samples were used and the experiments were executed twice on every sample.\\
Lastly, XP-spectroscopy was performed using VersaProbe II microscope (Physical Electronics GmbH) and CasaXPS software (Casa Software Ltd). The Dual-Beam charge neutralizer was driven using electrons with approximately 2 eV and Argon ions (Ar$^+$) with 10 eV. During measurements, the surface potential was about -2 V due to higher electron flux (whereas the S/S samples were grounded due to their conductibility). Radiation source was a monochromatic Al-k $\alpha$ radiator with an energy of 1,486 eV. The X-ray power was about 90 W and the analysis area was 1.3 mm x 100 $\mu m$ at high power mode. Transmitting energy of the analyzer was at 93.9 eV with a step size of 0.8 eV and an integration time of 50 ms. With the XPS microscope, one plasma-treated sample and one reference sample of every material were analyzed.
\section{Results}
\begin{figure}[htb!]
    
  \begin{subfigure}[b]{0.49\columnwidth}
      \centering
    \includegraphics[width=0.9\columnwidth]{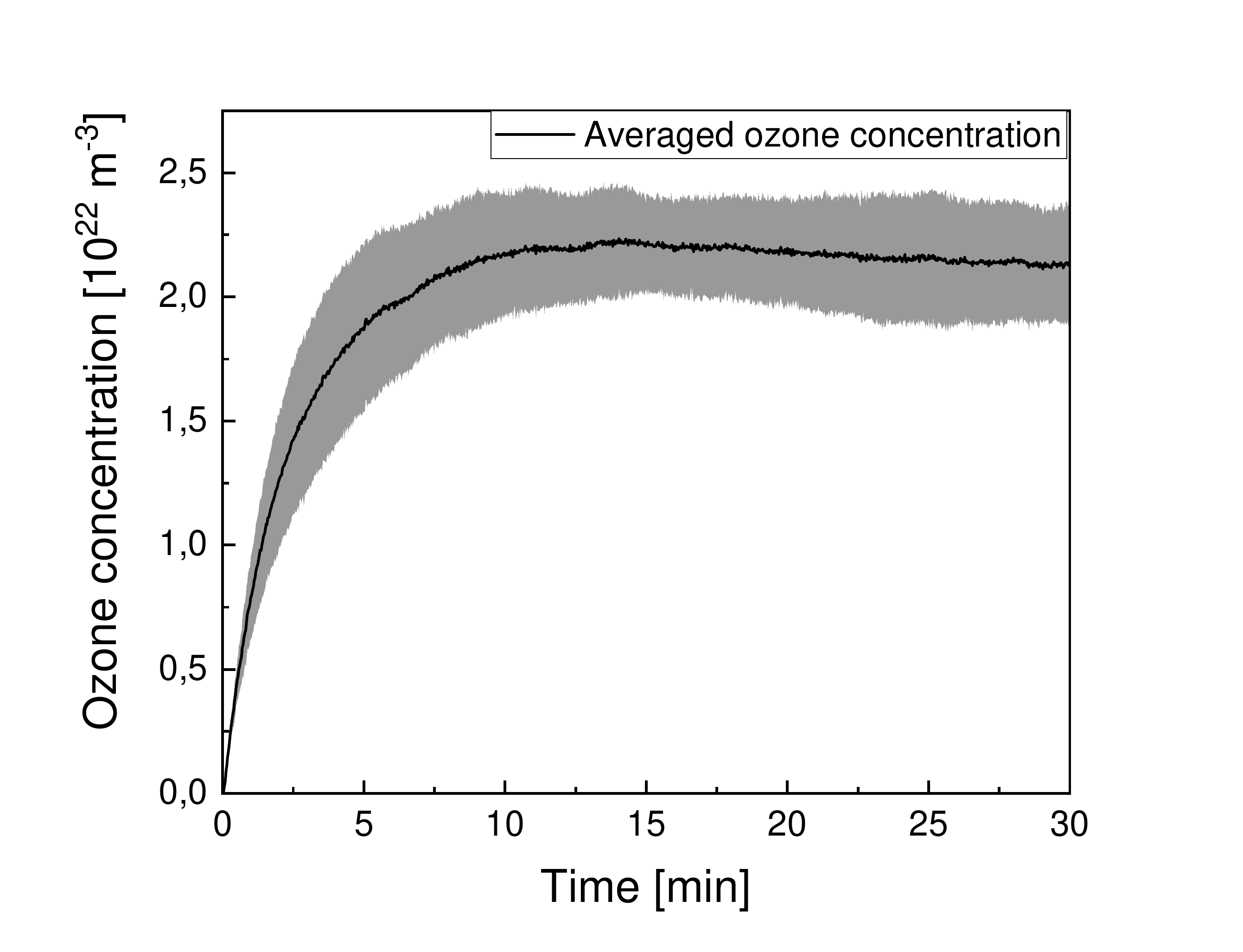}
        \caption{Ozone concentration, derived with Beer-Lambert-Law, during 30 minutes of plasma treatment averaged over three treatment cycles with 10 minutes breaks between each cycle. The gray areas show the standard deviation.}
         \label{ozon}
       \end{subfigure}%
        ~ 
        \begin{subfigure}[b]{0.49\columnwidth}
                \centering
          \includegraphics[width=0.9\columnwidth]{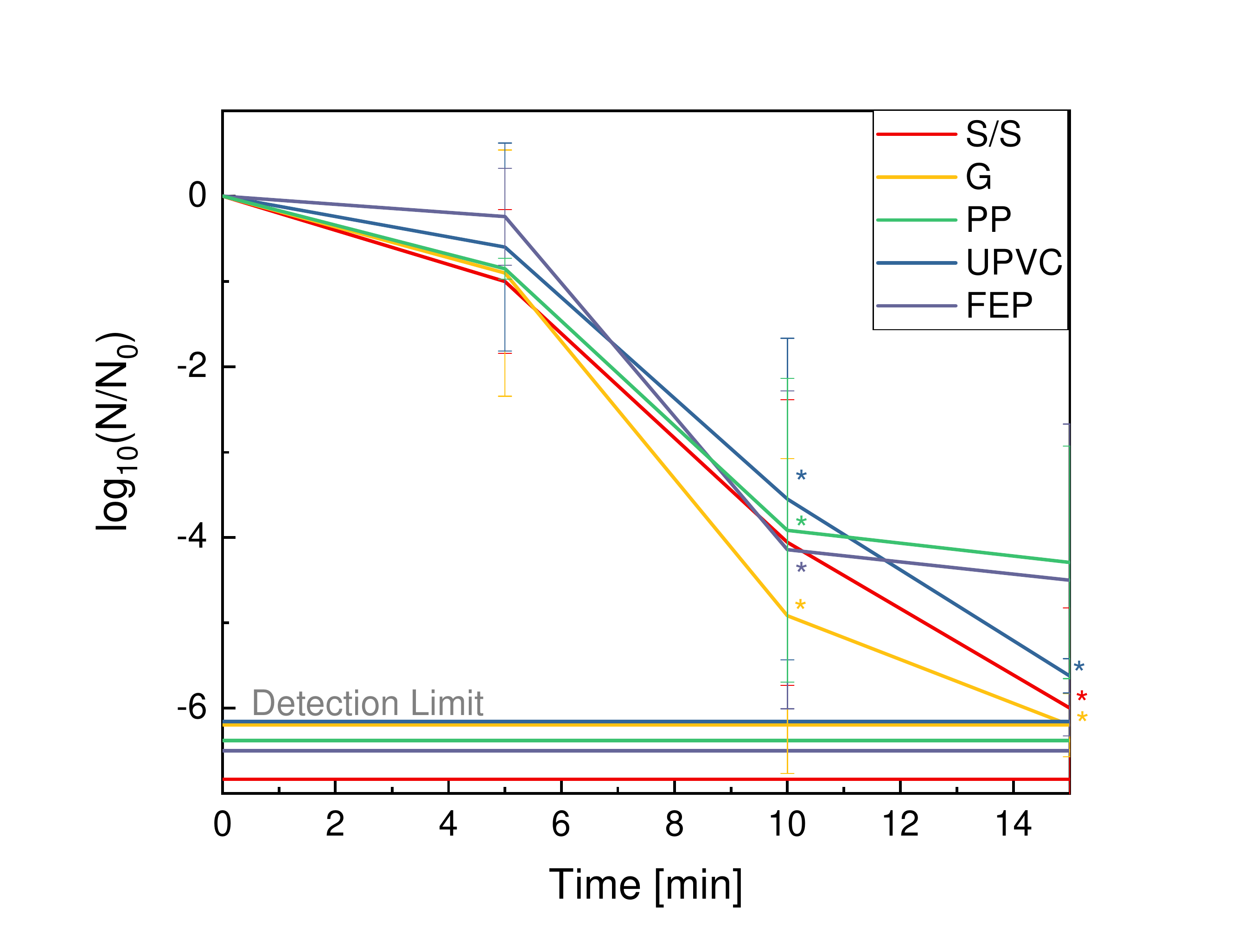}
              \caption{Log reduction of \textit{Bacillus atrophaeus} endospores at different treatment times. N is the number of CFU after plasma treatment, N$_0$ the number of CFU for the reference samples. Samples with no CFU were counted as explained in Materials\&Methods.}
          \label{Bakterien}
        \end{subfigure}
        \caption{Ozone concentration during plasma treatment time of 30 minutes (left) and log reduction curves of \textit{Bacillus atrophaeus} endospores on different materials (right).} \label{Bakterien&Ozon}
  \end{figure}

\begin{figure*}[htb!]
        \centering
        \begin{subfigure}[b]{0.49\columnwidth}
                \centering
                \includegraphics[width=0.9\columnwidth]{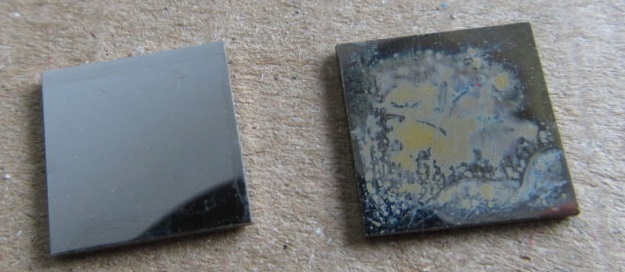}
                \subcaption{ S/S macroscopic; left is reference sample, right is plasma-treated}
        \end{subfigure}%
    ~
        \begin{subfigure}[b]{0.49\columnwidth}
                \centering
                \includegraphics[width=0.9\columnwidth]{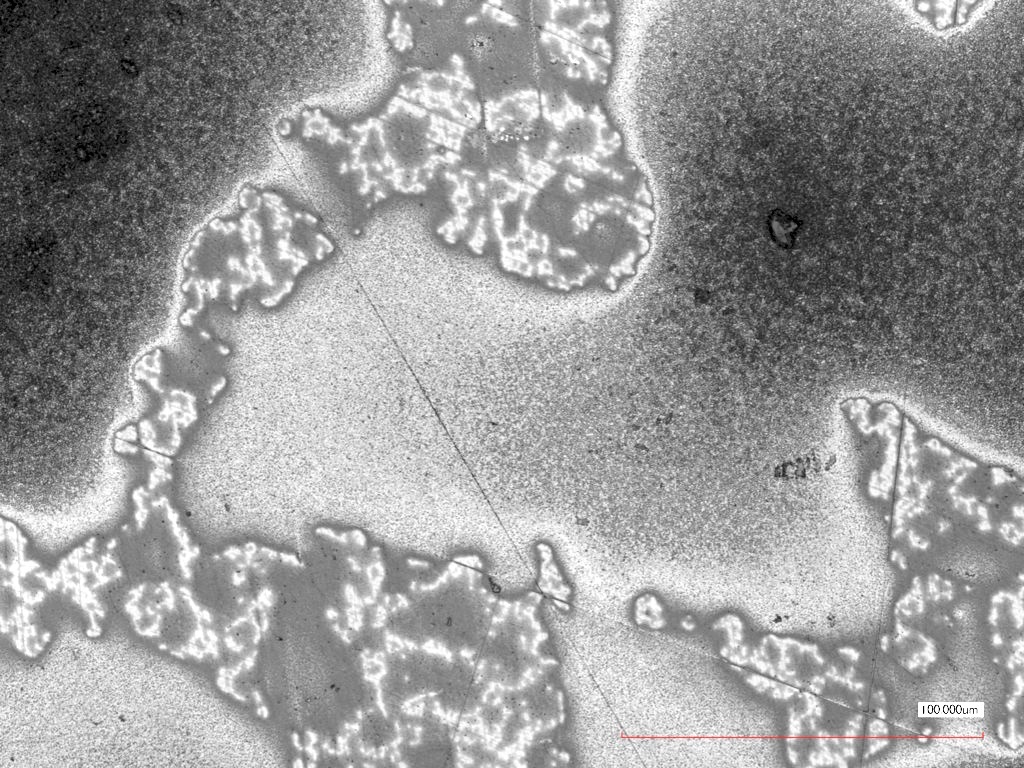}
                \subcaption{Laser microscopic S/S image (plasma-treated)}
        \end{subfigure}
            \caption{After plasma-treatment, a brown, rusty cover appeared on the S/S samples. It was visible to the naked eye (a). More structures could be found with laser microscopy images (b).}\label{rost}
  \end{figure*}

\begin{figure*}[htb!]
        \centering
        \begin{subfigure}[b]{0.49\columnwidth}
                \centering
                \includegraphics[width=0.9\columnwidth]{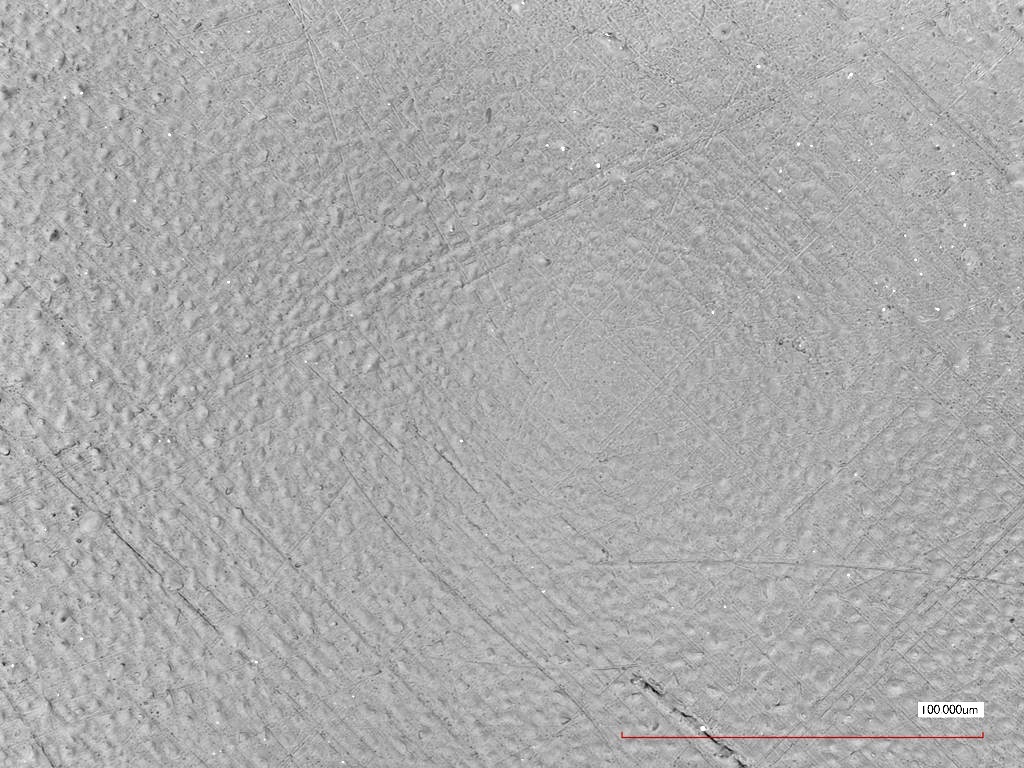}
                \subcaption{ PP plasma-treated}
        \end{subfigure}%
    ~
        \begin{subfigure}[b]{0.49\columnwidth}
                \centering
                \includegraphics[width=0.9\columnwidth]{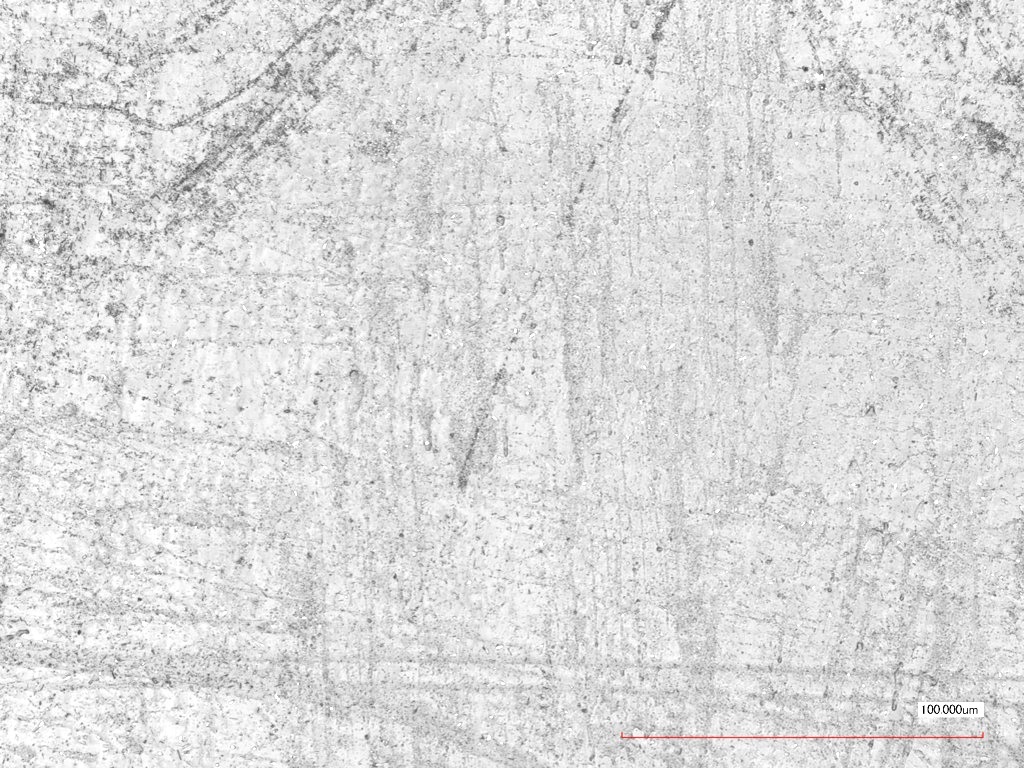}
                \subcaption{PP reference}
        \end{subfigure}
        
        \begin{subfigure}[b]{0.49\columnwidth}
                \centering
                \includegraphics[width=0.9\columnwidth]{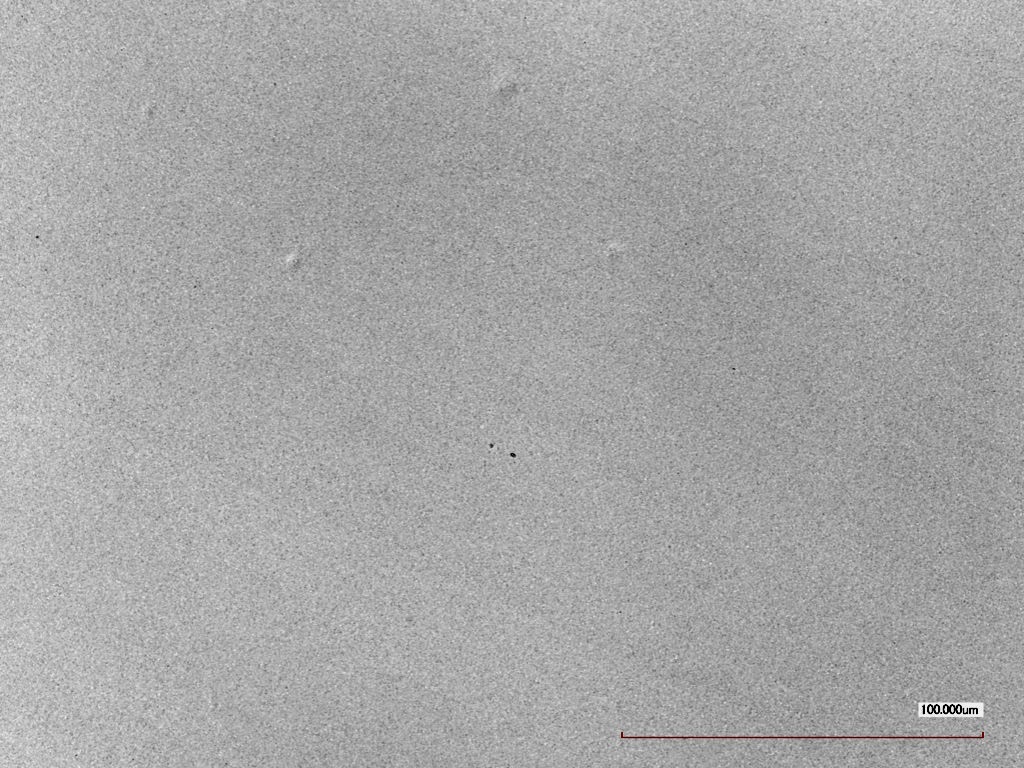}
                \subcaption{UPVC plasma-treated}
        \end{subfigure}%
    ~
   \begin{subfigure}[b]{0.49\columnwidth}
                \centering
                \includegraphics[width=0.9\columnwidth]{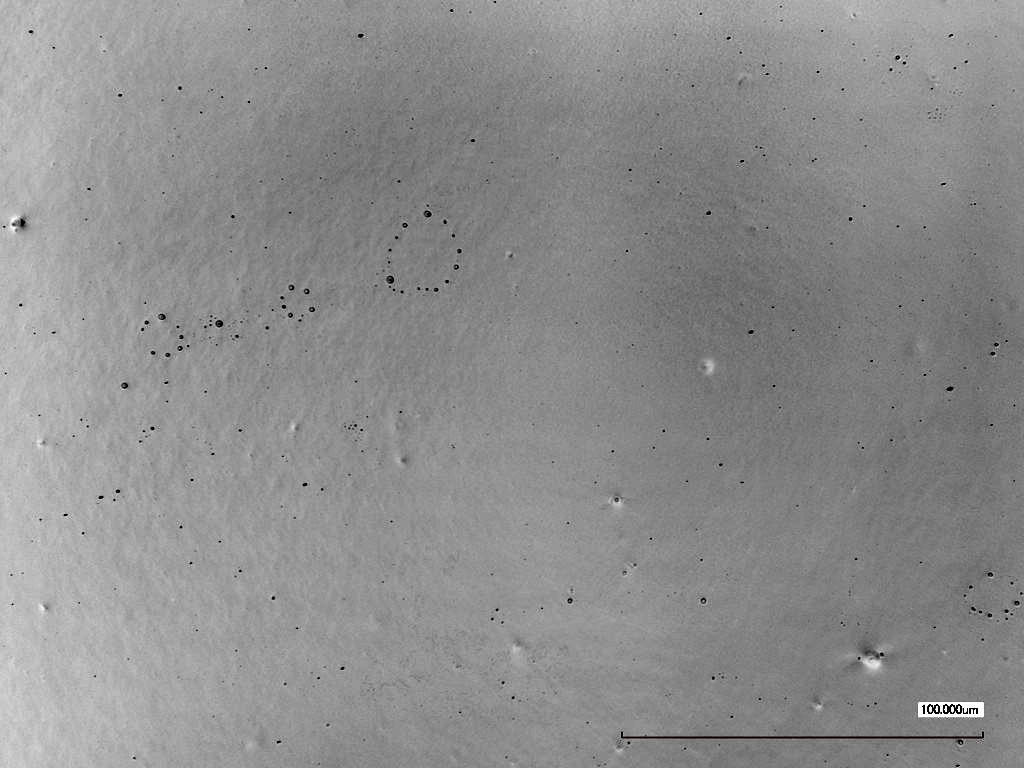}
                \subcaption{UPVC reference}
        \end{subfigure}
        \caption{Laser microscopy identified a modification of the samples' surfaces. While the reference sample of PP (b) was rather flat and even, the plasma-treated sample (a) had a lot of dints at its surface. In contrast, plasma-treated UPVC samples (c) had a very planar morphology, while reference samples showed circular structures (d).}\label{laserimages}
  \end{figure*}

\begin{figure*}[htb!]
        \centering
        \begin{subfigure}[b]{0.49\columnwidth}
                \centering
                \includegraphics[width=0.9\columnwidth]{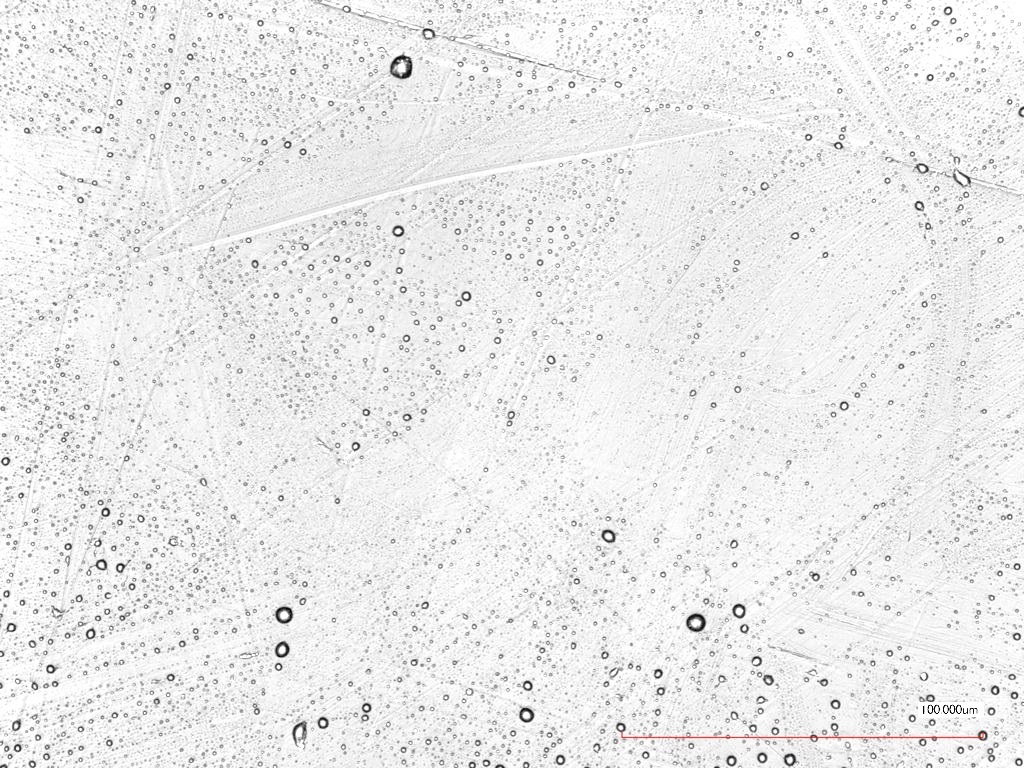}
                \subcaption{FEP plasma-treated}
        \end{subfigure}%
    ~
        \begin{subfigure}[b]{0.49\columnwidth}
                \centering
                \includegraphics[width=0.9\columnwidth]{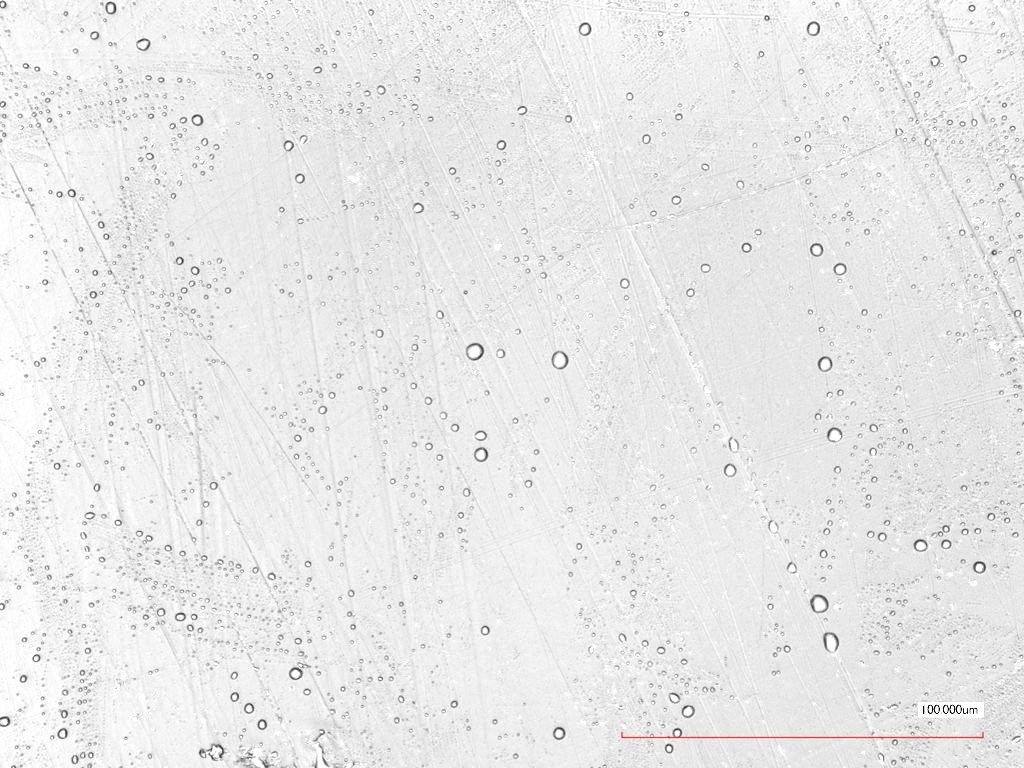}
                \subcaption{FEP reference}
        \end{subfigure}
        \caption{FEP samples showed no modifications beetween plasma-treated (a) and reference sample (b) under laser microscopy.}\label{feplaser}
  \end{figure*}

In this paper, surface modifications caused by long-term SMD plasma treatment on S/S, glass, PP, UPVC and FEP were investigated in detail using laser microscopy, contact angle measurements and XPS-quantification. Moreover, sporicidal properties of the SMD plasma were verified, treating \textit{Bacillus atrophaeus} endospores inoculated on the probed materials, and the ozone concentration in the plasma chamber was measured. As shown in figure \ref{ozon}, it takes approximately 5 min to reach a steady state ozone concentration. \\
\noindent Figure \ref{Bakterien} shows the reduction of \textit{Bacillus atrophaeus} endospores on different materials as a function of plasma treatment time at room temperature with a relative humidity of about 60-70\%.
\\ 
\noindent For S/S, a log reduction of 6.0 could be reached within 15 minutes of treatment time, for UPVC and glass log reduction was 6.2 and 5.6, respectively. For those materials spore reduction reaches close to the detection limits, whereas for FEP and PP a log reduction of 4.5 and 4.3, respectively, could be achieved. During the first 5 minutes, only a small bacteria reduction with a maximum of log 1 could be obtained.
At a time period between 5 and 10 minutes, the reduction function became steep. Consequently, D-values were between 5 (S/S) and 6 minutes (FEP), while the reduction from log 1 to log 2 took only between 1.0 (glass) and 2 minutes (UPVC). 
After 10 minutes, log reduction became slower.\\
\noindent The surface analysis of the material samples showed a wide spectrum of modifications due to SMD treatment.\\
The surface roughness of treated materials stayed approximately the same (table \ref{Ra}), only for S/S, significant changes were observed, and a rusty-colored covering, also visible to the naked eye, was found after plasma-treatment (figure \ref{rost}). Nevertheless, laser microscopy images of PP and UPVC showed a change of appearance (figure \ref{laserimages}) whereas there were no visible changes for FEP samples (figure \ref{feplaser}).

\begin{table*}[htb!]
\centering
\begin{tabular}{|c|c|c|}
\hline
               & \textbf{plasma-treated} & \textbf{reference}     \\ \hline
\textbf{S/S}   & 0.111$\pm$0.084$\mu m$  & 0.016$\pm$0.002$\mu m$ \\ \hline
\textbf{Glass} & -                       & -                      \\ \hline
\textbf{PP}    & 0.058$\pm$0.006$\mu m$  & 0.056$\pm$0.011$\mu m$ \\ \hline
\textbf{UPVC}  & 0.018$\pm$0.003$\mu m$  & 0.014$\pm$0.003$\mu m$ \\ \hline
\textbf{FEP}   & 0.037$\pm$0.006$\mu m$  & 0.036$\pm$0.010$\mu m$ \\ \hline
\end{tabular}\caption{Roughness value $R_a$ of different materials with and without plasma-treatment. Only S/S shows a significant change in surface roughness.}\label{Ra}
\end{table*}

\begin{figure*}[htb!]
        \centering
        \begin{subfigure}[b]{0.49\columnwidth}
                \centering
                \includegraphics[width=0.9\columnwidth]{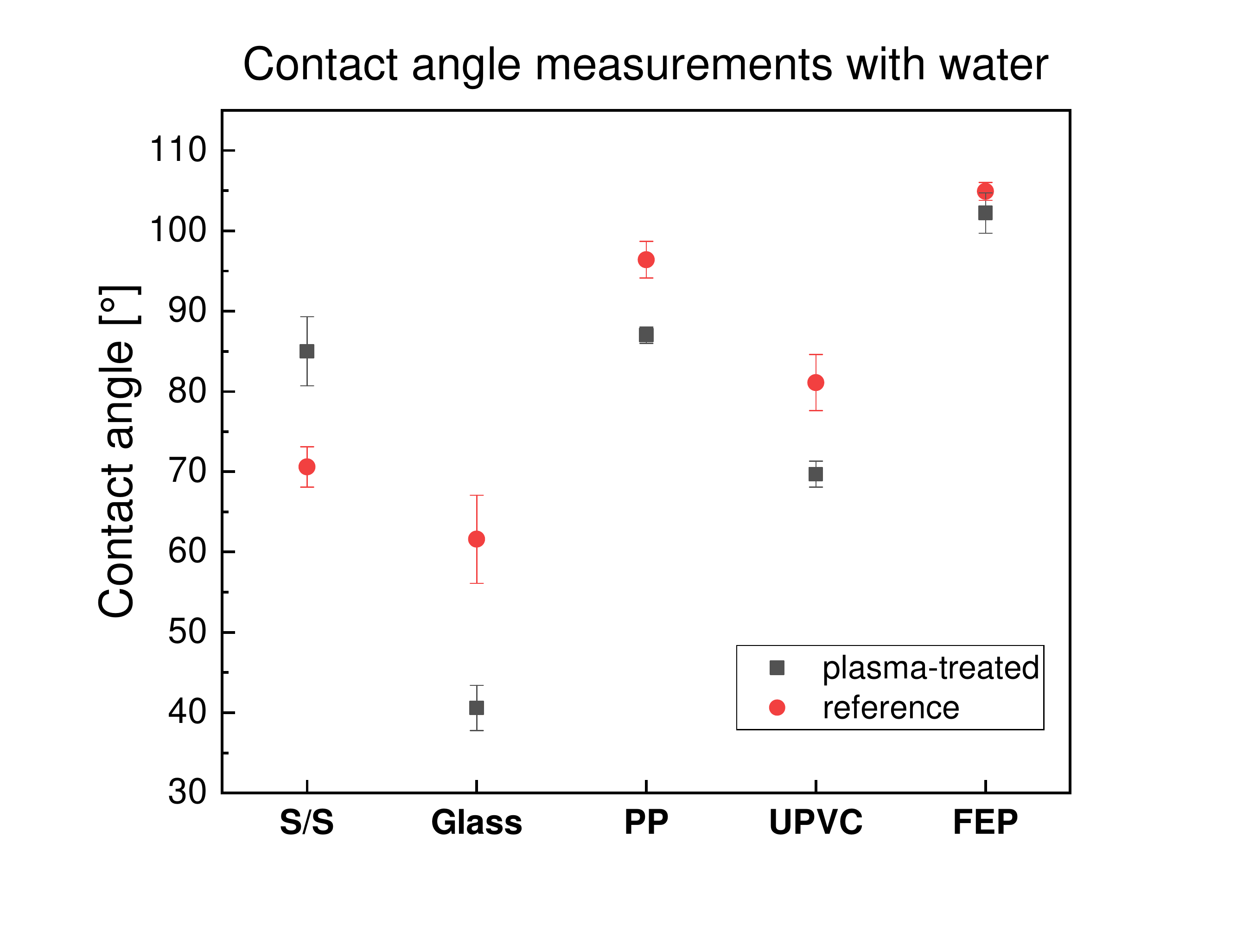}
                \subcaption{ }
        \end{subfigure}%
    ~
\begin{subfigure}[b]{0.49\columnwidth}
                \centering
                \includegraphics[width=0.9\columnwidth]{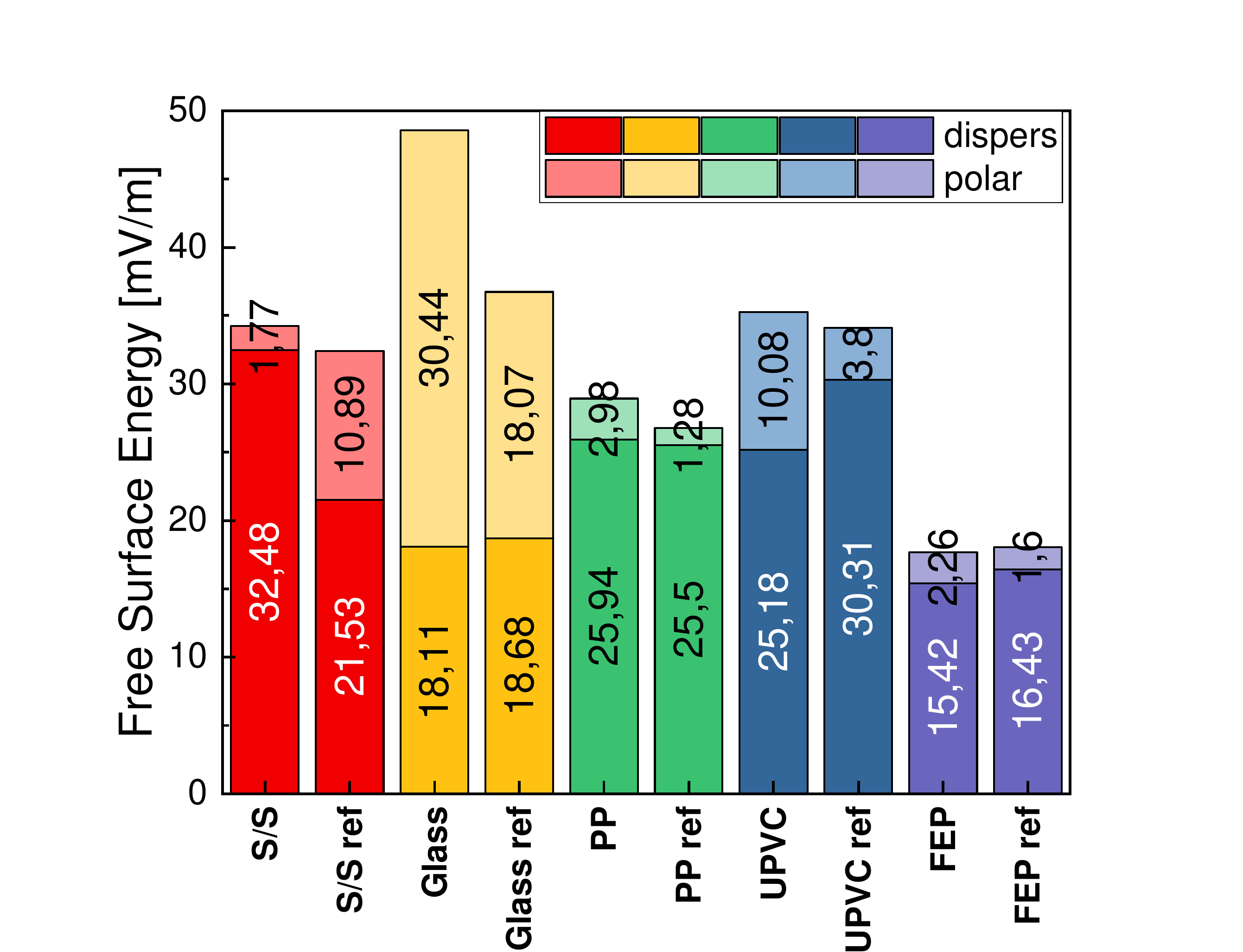}
                \subcaption{ }
        \end{subfigure}
        \caption{Results of contact angle measurements of different materials with deionized water (a) and calculated polar and disperse fractions of free surface energies of the materials (b), respectively.}\label{ContactAngle}
  \end{figure*}

\noindent In accordance with the laser microscopy results, changes of the contact angle between probed samples and reference liquids (deionized water, ethanol, diiodomethane and ethylene glycol) for plasma-treated materials were detected (e.g. figure \ref{ContactAngle} (a)). Again, FEP showed no detectable difference between plasma-treated and reference sample, in contrast to all other materials. Note, that only for S/S the contact angle got higher for the plasma-treated case, whereas glass, PP and UPVC became more hydrophilic (figure \ref{ContactAngle} (a)). Using OWRK-modell, the samples' free surface energies with their disperse and polar fractions were calculated (figure \ref{ContactAngle} (b)). S/S showed an increase of disperse fraction of free surface energy for plasma-treated case, while for glass, PP and UPVC the polar fraction increased. Besides, glass samples showed a significant increase of total free surface energy with a high increase of its polar fraction. In contrast, no significant changes between plasma-treated and reference sample for FEP could be detected, yet there was a slight increase in the polar fraction and a slight decrease in the disperse fraction.\\
  \\\noindent Finally, XPS analysis measured a quantified value for the surfaces' elemental composition. The results for plasma-treated FEP samples were again very similar to those of the reference samples (table \ref{XPS}). There was no nitrogen found and only small amounts of oxygen.\\
  However, glass, PP and UPVC showed an increase of oxygen and nitrogen species on the surface. Although there was no nitrogen found for reference samples, there was roughly 0.5-1\% found for plasma-treated samples.\\
  Again, S/S showed the greatest change of elemental composition with an augmentation of the atomic concentration of oxygen, nitrogen and ferric on the surface. Note, that oxygen and ferric concentration almost doubled while nitrogen concentration more than quintupled.
  \begin{table*}[]
  \centering
\begin{tabular}{|c||c|c|c|c||c|c|c|c|}
               \hline    & \multicolumn{4}{c||}{\textbf{plasma-treated [atomic-\%]}} & \multicolumn{4}{c|}{\textbf{reference [atomic-\%]}} \\
                \hline    & \textbf{O}                                                        & \textbf{C}      & \textbf{N}     & \textbf{Fe}    & \textbf{O}                                                   & \textbf{C}      & \textbf{N}     & \textbf{Fe}    \\
\hline\hline 
\textbf{S/S}            & 54.1\%                                                   & 32.0\% & 7.5\% & 3.6\% & 36.4\%                                              & 56.2\% & 1.3\% & 1.6\% \\
\hline \textbf{Glass}               & 67.0\%                                                   & 6.6\%  & 0.8\% &       & 54.5\%                                              & 15.5\% & 0.0\% &       \\
\hline \textbf{PP}                  & 10.8\%                                                   & 88.6\% & 0.6\% &       & 8.5\%                                               & 90.2\% & 0.0\% &       \\
\hline \textbf{UPVC}                & 10.7\%                                                   & 68.3\% & 0.7\% &       & 6.9\%                                               & 78.1\% & 0.0\% &       \\
\hline \textbf{FEP}                 & 1.6\%                                                    & 37.9\% & 0.0\% &       & 1.2\%                                               & 36.6\% & 0.0\% &     \\\hline 
\end{tabular}
\caption{Elemental composition of the material samples measured with XPS quantification with (left) and without (right) plasma treatment.} 
\label{XPS}
\end{table*}
\clearpage
\section{Discussion}
After treatment with SMD plasma, three different material reactions can be found. As a first group, there is FEP, where no significant surface changes are found. Although there were minor changes in contact angle measurements and therefore in FEP's free surface energy, the measured values were so close to each other, that they laid within the error range. Hence, these changes are negligible. In the second group, slight changes in surface appearance and atomic concentration, accompanied with higher polar fractions of free surface energy were found. These effects were found for glass, as well as the polymers PP and UPVC. \\
As third group, there is S/S with significantly changed elemental composition, free surface energy properties and surface appearance.\\
Ozone concentration measurements revealed, that ozone is one of the main plasma product produced during the experiments, since its concentration reaches a steady-state range after 5 minutes of treatment time. The plasma settings therefore generate an SMD plasma in the so-called \grqq ozone mode\grqq ~\cite{Tschang,ozonmode,ozone2}. Yet, nitrogen as major element in air also participates in plasma-surface interactions, as the XPS-measurements revealed.
Hence, two different mechanisms explain the appearance of the rusty crust found on S/S samples, identified as ferrous oxide:
Firstly, nitrogen and oxygen interact as follows, producing nitric acid. Nitric acid oxidizes ferrous iron to ferrous oxide, with nitrogen dioxide and water as side products:\\    
\begin{center}
$ N_2 + O_2 \rightarrow 2NO$\\
$4NO + 3O_2+2H_2O \rightarrow 4HNO_3 $   \\
$ 2HNO_3+Fe \rightarrow 2NO_2+H_2O+FeO$\\
\end{center}
\noindent A second mechanism describes the reaction of ferrous iron and ozone:\\
\begin{center}
$2Fe+O_3+3H_2O \rightarrow 2Fe(OH)_3$ \\
$2Fe(OH)_3 \rightarrow Fe_2O_3+3H_2O $
\end{center}
Note, that in both reaction mechanisms water, which was accessible during the plasma treatment due to air humidity, served as a catalyst. The importance of high air humidity was also shown in \cite{Sandra}.
These findings support the findings of Shimizu \textit{et al.}\cite{aluminium}, where aluminum was found to oxidize after exposure to CAP. The slight changes in surface appearance of PP and UPVC can be explained with etching mechanisms as they were found in \cite{etching}, yet they were minor because of the weaker surface interaction of afterglow, that was used in this setup \cite{afterglowvorteil}. \\
Sporicidal experiments showed no parallels to the found surface modification groups because sporicidal effects were strongest for glass and S/S, followed by UPVC, and weakest for FEP and PP. Therefore, it is more likely, that differences in sporicidal effects are due to their different hydrophobic behavior and layering. If one compares the material's contact angles with water (figure \ref{ContactAngle}) and the images of sample preparation (figure \ref{dry}), it becomes obvious that the size of the area containing dried master solution in the experiments was smaller for PP and FEP because of their hydrophobic properties, whereas UPVC, glass and S/S were more hydrophilic, thus the same number of endospores were spread over a larger surface. As a consequence, layering phenomena, that shield the probed endospores partly, are more likely to influence the results for FEP and PP samples. In \cite{Sandra} and \cite{layer}, the link between layering mechanism and reduced bactericidal properties was also described.\\

\section{Conclusion}
After 16 hours of SMD plasma treatment, three reaction groups of the different materials, S/S, UPVC, PP, FEP and glass, were found.\\
While the surface of FEP samples showed no reaction to plasma treatment, the other material's surfaces changed. These changes were found in XP spectroscopy (attachment of oxygen and nitrogen), contact angle measurements and free surface energies (change in hydrophobia, change of polar and disperse fractions of free surface energy), as well as in laser microscopy (slight changes of surface appearance for UPVC and PP and a higher surface roughness and a rusty crust for S/S). Especially for S/S, the ferrous iron in the alloy was found to react with reactive nitrogen and oxygen species and combine to iron oxide. Consequently, the use of CAP in medical applications has to be considered carefully. The choice of material influences the plasma-surface reaction crucially, and advantages, e.g. low-temperature and difficult-geometry sterilization, and disadvantages, such as etching mechanisms of the surface and chemical reactions of treated materials with the surrounding air, have to be taken into consideration. In future research, the change of surface properties due to CAP treatment for more materials has to be done, especially for more polymers and different stainless steel alloys. Moreover, in future research more plasma parameters have to be investigated to find the optimal settings for a surface-preserving CAP treatment.\\
Sporicidal properties on the other hand were found to be only dependent on the materials' hydrophobic properties, and hence layering phenomena, but no correlation between the surfaces' sensitivity to CAP treatment and sporicidal effects was found. Nevertheless, the number of endospores on the sample's area, that is needed to prove sterilization properties, does not correlate with realistic cases, because rinsing and other cleaning methods reduce the number of microorganisms before sterilization treatment, and microorganisms are usually spread over a larger surface. In future work the aim will be to use a larger surface area with multiple droplets with lower concentrations spread over it to achieve higher spore surface density exposed to plasma.\\
\noindent The aim of this study was the identification of possible surface modifications of different materials due to long-term sporicidal cold atmospheric plasma treatment. Changes in appearance, free surface energy and surface elemental composition were found after 16 hours of plasma treatment for most materials, while bactericidal effects of the experimental setup were proven.
\section{Acknowledgments} S. Moritz wants to thank her collaborators Janosch J. Perlbach and Markus G\"ottlicher for their support in the laboratory.
\section*{References}
\bibliography{literatur}

\begin{thebibliography}{10}

\bibitem{obst}
Critzer FJ, Kelly-Wintenberg K, South SL, Golden DA.
\newblock Atmospheric plasma inactivation of foodborne pathogens on fresh
  produce surfaces.
\newblock Journal of food protection. 2007;70(10):2290--2296.

\bibitem{tissue}
Stoffels E, Sakiyama Y, Graves DB.
\newblock Cold atmospheric plasma: charged species and their interactions with
  cells and tissues.
\newblock IEEE Transactions on Plasma Science. 2008;36(4):1441--1457.

\bibitem{Sandra}
Mandler J, Moritz S, Binder S, Shimizu T, M{\"u}ller M, Thoma M, et~al.
\newblock Disinfection of dental equipment—Inactivation of Enterococcus
  mundtii on stainless steel and dental handpieces using surface
  micro-discharge plasma.
\newblock Plasma Medicine. 2017;7(4).

\bibitem{zahn1}
Arora V, Nikhil V, Suri N, Arora P.
\newblock Cold atmospheric plasma (CAP) in dentistry.
\newblock Dentistry. 2014;4(1):1.

\bibitem{zahn2}
{Jiang} C, {Chen} M, {Schaudinn} C, {Gorur} A, {Vernier} PT, {Costerton} JW,
  et~al.
\newblock Pulsed Atmospheric-Pressure Cold Plasma for Endodontic Disinfection
  $^{\ast}$.
\newblock IEEE Transactions on Plasma Science. 2009 July;37(7):1190--1195.

\bibitem{sterilization}
Lee K, Paek Kh, Ju WT, Lee Y.
\newblock Sterilization of bacteria, yeast, and bacterial endospores by
  atmospheric-pressure cold plasma using helium and oxygen.
\newblock Journal of microbiology (Seoul, Korea). 2006;44(3):269--275.

\bibitem{spores}
Kl{\"a}mpfl TG, Isbary G, Shimizu T, Li YF, Zimmermann JL, Stolz W, et~al.
\newblock Cold atmospheric air plasma sterilization against spores and other
  microorganisms of clinical interest.
\newblock Appl Environ Microbiol. 2012;78(15):5077--5082.

\bibitem{infection}
Nguyen L, Lu P, Boehm D, Bourke P, Gilmore BF, Hickok NJ, et~al.
\newblock Cold atmospheric plasma is a viable solution for treating orthopedic
  infection: a review.
\newblock Biological chemistry. 2018;400(1):77--86.

\bibitem{Tschang}
Tschang CYT, Thoma M.
\newblock In vitro comparison of direct plasma treatment and plasma activated
  water on Escherichia coli using a surface micro-discharge.
\newblock Journal of Physics D: Applied Physics. 2019 nov;53(5):055201.
\newblock Available from: \url{https://doi.org/10.1088%2F1361-6463%2Fab522a}.

\bibitem{rons}
Kim SJ, Chung T.
\newblock Cold atmospheric plasma jet-generated RONS and their selective
  effects on normal and carcinoma cells.
\newblock Scientific reports. 2016;6:20332.

\bibitem{etching}
Kuzminova A, Kretkov{\'a} T, Kyli{\'a}n O, Hanu{\v{s}} J, Khalakhan I, Prukner
  V, et~al.
\newblock Etching of polymers, proteins and bacterial spores by atmospheric
  pressure DBD plasma in air.
\newblock Journal of Physics D: Applied Physics. 2017;50(13):135201.

\bibitem{aluminium}
Shimizu S, Barczyk S, Rettberg P, Shimizu T, Klaempfl T, Zimmermann JL, et~al.
\newblock Cold atmospheric plasma--A new technology for spacecraft component
  decontamination.
\newblock Planetary and Space Science. 2014;90:60--71.

\bibitem{walsh}
Walsh JL, Shi J, Kong MG.
\newblock Contrasting characteristics of pulsed and sinusoidal cold atmospheric
  plasma jets.
\newblock Applied Physics Letters. 2006;88(17):171501.

\bibitem{fridman}
Fridman G, Shereshevsky A, Jost MM, Brooks AD, Fridman A, Gutsol A, et~al.
\newblock Floating electrode dielectric barrier discharge plasma in air
  promoting apoptotic behavior in melanoma skin cancer cell lines.
\newblock Plasma Chemistry and Plasma Processing. 2007;27(2):163--176.

\bibitem{morfill}
Morfill G, Kong MG, Zimmermann J.
\newblock Focus on plasma medicine.
\newblock New Journal of Physics. 2009;11(11):115011.

\bibitem{klaempfl}
Maisch T, Shimizu T, Isbary G, Heinlin J, Karrer S, Kl\"ampfl T, et~al.
\newblock Contact-Free Inactivation of Candida albicans Biofilms by Cold
  Atmospheric Air Plasma.
\newblock Applied and Environmental Microbiology. 2012;78(12):4242--4247.
\newblock Available from: \url{http://aem.asm.org/content/78/12/4242.abstract}.

\bibitem{gas}
Schmidt-Bleker A, Winter J, B{\"o}sel A, Reuter S, Weltmann KD.
\newblock On the plasma chemistry of a cold atmospheric argon plasma jet with
  shielding gas device.
\newblock Plasma Sources Science and Technology. 2015;25(1):015005.

\bibitem{hybrid}
Ly L, Jones S, Shashurin A, Zhuang T, Rowe W, Cheng X, et~al.
\newblock A new cold plasma jet: Performance evaluation of cold plasma, hybrid
  plasma and argon plasma coagulation.
\newblock Plasma. 2018;1(1):189--200.

\bibitem{helium1}
Darny T, Pouvesle JM, Fontane J, Joly L, Dozias S, Robert E.
\newblock Plasma action on helium flow in cold atmospheric pressure plasma jet
  experiments.
\newblock Plasma Sources Science and Technology. 2017;26(10):105001.

\bibitem{helium2}
Lin L, Lyu Y, Trink B, Canady J, Keidar M.
\newblock Cold atmospheric helium plasma jet in humid air environment.
\newblock Journal of Applied Physics. 2019;125(15):153301.

\bibitem{afterglow}
Mok C, Lee T, Puligundla P.
\newblock Afterglow corona discharge air plasma (ACDAP) for inactivation of
  common food-borne pathogens.
\newblock Food Research International. 2015;69:418--423.

\bibitem{afterglowvorteil}
Moreau S, Moisan M, Tabrizian M, Barbeau J, Pelletier J, Ricard A, et~al.
\newblock Using the flowing afterglow of a plasma to inactivate Bacillus
  subtilis spores: Influence of the operating conditions.
\newblock Journal of Applied Physics. 2000;88(2):1166--1174.

\bibitem{powerdensity}
Ho{\l}ub M.
\newblock On the measurement of plasma power in atmospheric pressure DBD plasma
  reactors.
\newblock International Journal of Applied Electromagnetics and Mechanics.
  2012;39(1-4):81--87.

\bibitem{sakiyama}
Sakiyama Y, Graves DB, Chang HW, Shimizu T, Morfill GE.
\newblock Plasma chemistry model of surface microdischarge in humid air and
  dynamics of reactive neutral species.
\newblock Journal of Physics D: Applied Physics. 2012;45(42):425201.

\bibitem{beer}
Bass A, Paur R.
\newblock The ultraviolet cross-sections of ozone: I. The measurements.
\newblock In: Atmospheric ozone. Springer; 1985. p. 606--610.

\bibitem{SessileDrop}
Staicopolus D.
\newblock The computation of surface tension and of contact angle by the
  sessile-drop method.
\newblock Journal of Colloid Science. 1962;17(5):439--447.

\bibitem{strom}
Str{\"o}m G, Fredriksson M, Stenius P.
\newblock Contact angles, work of adhesion, and interfacial tensions at a
  dissolving hydrocarbon surface.
\newblock Journal of colloid and interface science. 1987;119(2):352--361.

\bibitem{ohm}
Ohm A, Lippold B.
\newblock Charakterisierung der Benetzbarkeit von Arzneistoffpulvern mit Hilfe
  der Sessile Drop-Technik. II: Kritische Oberfl{\"a}chenspannung und
  Randwinkel/Oberfl{\"a}chenspannungs-Kurven.
\newblock Pharmazeutische Industrie. 1986;48(5):508--513.

\bibitem{o}
Owens DK, Wendt R.
\newblock Estimation of the surface free energy of polymers.
\newblock Journal of applied polymer science. 1969;13(8):1741--1747.

\bibitem{r}
Rabel W.
\newblock Einige Aspekte der Benetzungstheorie und ihre Anwendung auf die
  Untersuchung und Ver{\"a}nderung der Oberfl{\"a}cheneigenschaften von
  Polymeren.
\newblock Farbe und Lack. 1971;77(10):997--1005.

\bibitem{k}
Kaelble D.
\newblock Dispersion-polar surface tension properties of organic solids.
\newblock The Journal of Adhesion. 1970;2(2):66--81.

\bibitem{ozonmode}
Pavlovich MJ, Chang HW, Sakiyama Y, Clark DS, Graves DB.
\newblock Ozone correlates with antibacterial effects from indirect air
  dielectric barrier discharge treatment of water.
\newblock Journal of Physics D: Applied Physics. 2013;46(14):145202.

\bibitem{ozone2}
Shimizu T, Sakiyama Y, Graves DB, Zimmermann JL, Morfill GE.
\newblock The dynamics of ozone generation and mode transition in air surface
  micro-discharge plasma at atmospheric pressure.
\newblock New Journal of Physics. 2012;14(10):103028.

\bibitem{layer}
Yu H, Perni S, Shi JJ, Wang DZ, Kong MG, Shama G.
\newblock Effects of cell surface loading and phase of growth in cold
  atmospheric gas plasma inactivation of Escherichia coli K12.
\newblock Journal of Applied Microbiology. 2006;101(6):1323--1330.
\newblock Available from:
  \url{https://sfamjournals.onlinelibrary.wiley.com/doi/abs/10.1111/j.1365-2672.2006.03033.x}.

\end{thebibliography}
\bibliographystyle{vancouver}
\end{document}